\documentclass[twocolumn,aps,superscriptaddress]{revtex4-1}
\usepackage{palatino}
\usepackage[latin1]{inputenc}
\usepackage{epsf}
\usepackage{amsmath,amssymb}
\usepackage{latexsym}
\usepackage{calc}
\usepackage{color}
\usepackage{shadow}
\usepackage{epsfig}

\newcommand{\ben}{\begin{equation}}
\newcommand{\een}{\end{equation}}
\newcommand{\bea}{\begin{eqnarray}}
\newcommand{\eea}{\end{eqnarray}}

\def\sss{\scriptscriptstyle\rm}

\def\c{_{\sss C}}
\def\s{_{\sss S}}
\def\^s{^{\sss S}}
\def\xc{_{\sss XC}}
\def\Hx{_{\sss HX}}
\def\Hxc{_{\sss HXC}}

\def\H{_{\sss H}}

\def\ext{_{\rm ext}}

\def\br{{\bf r}}

\begin{document}

\title{Challenging Adiabatic Time-dependent Density Functional Theory with a Hubbard Dimer: The Case of Time-Resolved Long-Range Charge Transfer}
\author{Johanna I. Fuks}
\author{Neepa T. Maitra}
\affiliation{Department of Physics and Astronomy, Hunter College and the City University of New York, 695 Park Avenue, New York, New York 10065, USA}
\date{\today}

\begin{abstract}
We explore an asymmetric two-fermion Hubbard dimer to test the accuracy of the
adiabatic approximation of time-dependent density functional theory in
modelling time-resolved charge transfer. We show that the model shares essential features of a ground state long-range molecule in real-space, and by applying a resonant field we show that the model also reproduces essential traits of the CT dynamics.
The simplicity of the model
allows us to propagate with an ``adiabatically-exact'' approximation,
i.e. one that uses the exact ground-state exchange-correlation
functional, and compare with the exact propagation.   This allows us to study the impact of the
time-dependent charge-transfer step feature in the exact correlation
potential of real molecules on the resulting dynamics. 
Tuning the parameters of the dimer allows a study both of charge-transfer between open-shell fragments and between closed-shell fragments. We find that the adiabatically-exact functional is unable
to properly transfer charge, even in 
situations where the adiabatically-exact resonance frequency is remarkably close to the exact resonance, and we analyze why.

\end{abstract}
\maketitle

\section{Introduction}
The transfer of an electron across a molecule is an essential process
in biology, chemistry, and physics, that needs to be accurately
described in order to computationally model phenomena in many topical
applications, e.g. photovoltaics, vision, photosynthesis, molecular
electronics, and the control of coupled electron-ion dynamics by
strong lasers(e.g. Refs~\cite{DP07,Rozzi13,Jailaubekov12,TTRF08,NR03,Polli,Sansone10}).
For most of these applications, a time-resolved picture of the charge
transfer (CT) is extremely useful, and often necessary, as has been
stressed in recent work, and the correlation between electrons as well
as between electrons and ions play a crucial role~\cite{SAMYG13}.  The
systems are large enough that time-dependent density functional theory
(TDDFT) is the only calculationally feasible
approach~\cite{RG84,TDDFTbook,Carstenbook}. It is well-known that the
standard functional approximations considerably underestimate CT
excitations, and there has been intense development of improved
functionals for this; in particular the optimally tuned hybrids
present a useful non-empirical approach~\cite{SKB09p}. However the
transfer of one electron from one region of space to another is
clearly a non-perturbative process and calls for calculations that go
beyond linear response and excitations. The success of TDDFT to date rests on its performance in the linear regime, however the theory applies also to dynamics far from equilibrium. 
 The performance of
functionals for CT in this regime paints a more hazy picture: there
have been calculations in good agreement with
experiment~(e.g. Ref. \cite{Rozzi13}) but failures have been reported too \cite{Nest}. It would be fair to say that it is not
always clear to what accuracy the TDDFT results can be trusted. Part
of the problem is that there are very few alternate practical
computational methods for correlated electronic dynamics to test
against.  Calculations on simplified model systems that can be solved
exactly, e.g. two-electron systems in one-dimension, have highlighted
prominent features that the approximate functionals lack, not just for CT
dynamics~\cite{FERM13}, but also more generally in the non-linear
regime~\cite{TGK08,RB09,FHTR11,EFRM12}. The errors that result from
the lack of these features appear to be sometimes very significant, and other
times less so.

TDDFT in practise is almost always synonomous with adiabatic TDDFT,
certainly in the non-linear regime. That is, the Kohn-Sham (KS) system
is propagated using an adiabatic exchange-correlation potential, where
the evolving density at time $t$ is input into a ground-state (gs)
functional: $v\xc^{\rm A}[n;\Psi_0,\Phi_0](\br,t) = v\xc^{\rm
  gs}[n(t)]({\bf r})$. 
There are two distinct sources of error in such
an approximation: one is from the choice of the gs functional
approximation, while the other is the adiabatic approximation itself.
To separate these the {\it adiabatically-exact} (AE)
approximation~\cite{TGK08} is defined: the instantaneous density is
input into the exact gs functional, $v\xc^{\rm
  AE}[n;\Psi_0,\Phi_0](\br,t)=v\xc^{\rm AE}[n](\br,t) = v\xc^{\rm
  exact\;gs}[n(t)]({\bf r})$.  This approximation neglects
memory-dependence that the exact functional is known to possess
(dependence on the density's history and true and KS initial states
$\Psi_0$ and $\Phi_0$) but is fully non-local in space, and, if the
true and KS states at time $t$ were actually gs's of some potential,
it would be exact at time $t$.

Since the exact gs exchange-correlation functional is not known, even
for one-dimensional two electron systems, $v\xc^{\rm AE}[n](\br,t)$
must be found via a numerical scheme, of an inverse problem type. A handful of
papers~\cite{TGK08,EFRM12,FERM13} have found $v\xc^{\rm AE}[n](\br,t)$
using an iterative scheme for some model systems: The exact density $n(t)$, found by
solving the interacting Schr\"odinger equation, provides the input to
an iterative procedure that finds at each $t$ of interest the
interacting and non-interacting gs's of density $n(t)$, along with the
potential in which they are the gs. Then, $v\xc^{\rm AE}[n(t)]({\bf r}) = v\ext^{\rm  exact\; gs}[n(t)]({\bf r}) - v\s^{\rm exact\; gs}[n(t)]({\bf r}) -
v\H[n(t)](\br)$ where $v\H[n](\br)$ is the electrostatic Hartree
potential. In most cases studied so far the AE potential $v\xc^{\rm AE}[n](\br,t)$ has been 
evaluated on the exact density $n(t)$, and compared with the exact
(memory-dependent) exchange-correlation potential
$v\xc[n,\Psi_0,\Phi_0](\br,t)$ at that time, to analyze how good the
AE approximation is, what features of the exact potential are missing,
etc.  
In one case, the AE potential was used to self-consistently propagate the KS orbitals, using at each time-step,  the AE potential
evaluated on the self-consistent instantaneous density.   Such a propagation provides a more useful assessment of the accuracy of the AE, as it measures directly the impact of the AE on the resulting dynamics. For example, it is possible that some features that might make the AE potential look significantly different than the exact, may in fact have a limited effect on the propagation. However, self-consistent AE propagation clearly
requires much more numerical effort, as many iterations need to be
performed at every time-step to find the potential to propagate in,
and it has only been done in a few examples \cite{TGK08, TK09b, RP10} in one-dimensional model systems. In regions where the density becomes too small, the inversion becomes unstable and noisy. 

In particular, for CT dynamics it is particularly challenging to
converge the iterative density-inversion scheme due to the very low
density region between the atoms. Yet, such a calculation is of great
interest for CT dynamics: not only because of its significance in the
phenomena mentioned earlier, but also because it is known the exact
functional develops features that the usual approximations lack. 
Ref.~\cite{FERM13} showed that for a two-electron model
molecule composed of closed-shell atoms and driven at the CT resonance, a step associated with the CT process gradually builds up over time in the exact correlation potential. 
A dynamical oscillatory step is superimposed on this (see Refs.~\cite{EFRM12,LFSEM13}), and is a generic feature of non-linear dynamics, not only in CT dynamics, that has a non-adiabatic  density-dependence.
The AE approximation fails to capture
the dynamical step but, when evaluated on the {\it exact} density, does yield a CT step although of a smaller size than the exact.
Such steps require functionals with a spatially non-local dependence on the density. 
The available approximations do not yield any step
structure whatsover: the dismal failure of ALDA, ASIC-LDA, and AEXX, none of which contain any step in the correlation potential, to transfer any charge
was shown (Fig 3 of Ref.~\cite{FERM13}) and attributed to this lack of
step structure.  We expect some blame must go to the adiabatic
approximation itself, but a question arises: is the partial step of
the AE approximation enough to give a reasonable description of the CT dynamics?  If  yes, this would greatly simplify the
on-going search for accurate functionals for non-perturbative CT: it would mean that one does need to build in spatial non-local density-dependence into the correlation functional approximation, but that one could get away with a time-local, i.e. adiabatic approximation. 
  To answer the question, we would need to propagate with
the AE self-consistently, but as discussed above, this procedure is numerically very challenging for CT dynamics. 
In a recent short paper~\cite{FM14a}, we have shown that the answer is no, by studying CT dynamics
in a two-fermion asymmetric Hubbard dimer, which shares the essential
features of CT dynamics with real-space molecules. 
Due to the small Hilbert space of the dimer the exact 
gs functional can be found via a constrained search, and then used in $v\xc^{\rm AE}(t)$ to self-consistently propagate the system. 
No iterative scheme is needed because the exact functional form of the gs Hartree-exchange-correlation (HXC) potential is known. 
This enabled us to 
assess errors  in the
adiabatic approximation for CT dynamics independently of those resulting from errors in the gs approximation used.
In this paper we give more details on the dimer model, and the procedure followed. 
Like in Ref.~\cite{FM14a}, we study both the cases of resonant CT
between closed-shells and between open-shells, by tuning the
potential-difference between the two sites. However, unlike
Ref~\cite{FM14a}, we choose this asymmetry such that the CT state of
the first case has a very similar density as the gs of the second, and
vice-versa.  Although this choice leads to the exact density-dynamics
in one case being a time-reversed version of the dynamics in the other
case, we find the AE dynamics does not have this property. The AE
approximation in the closed-shell case is better for longer than for
the open-shell case, where it fails almost immediately; yet in either case, it fails to properly transfer the charge.  
The hope that the step seen in the AE approximation evaluated on the
exact density, albeit smaller than the exact, is enough to do a
reasonable job for CT processes is dashed.  A further result is an
expression for the interacting frequencies of the system in terms of
the KS ones and the HXC kernel. Using this, we compare the exact
resonant frequency of the interacting system with that predicted by
the AE approximation.

In Section~\ref{sec:model} we introduce the
model, its ground-state energies and potentials, and the exact
time-dependent KS potential. In Section~\ref{sec:cs--cs}, we present
the parameters used to study CT between closed-shells, give details of
the eigenstates of the interacting and KS systems, and propagate the
system with a resonant field to induce Rabi oscillations between the
ground and CT excited state. We compare the exact propagation with
that resulting from the AE propagation and discuss features of the
potentials. Section~\ref{sec:os--os} contains the analogous analysis
for the case of CT between open-shells. In Section~\ref{sec:AEres}
we derive a formula for the interacting frequencies of the system in
terms of the KS ones and the HXC kernel.  This is used to find the AE
resonant frequency, and compare with the exact in each case.

\section{The model}
\label{sec:model}
The Hamiltonian of the
two-site interacting Hubbard model with on-site repulsion $U$ and hopping 
parameter $T$ \cite{AG02,LU08,V08,CF12,FT12,FFTAKR13,B08,FM14a,CC13} reads:
\begin{align}
 \hat{H}= & -T \sum_\sigma \left( \hat{c}_{L\sigma}^\dag \hat{c}_{R\sigma} +\hat{c}_{R\sigma}^\dag\hat{c}_{L\sigma} \right)
+ U \left( \hat{n}_{L \uparrow} \hat{n}_{L \downarrow} + \hat{n}_{R \uparrow} \hat{n}_{R \downarrow}\right)  \nonumber \\ 
          & +  \frac{\Delta v (t)}{2}(\hat{n}_{L} -\hat{n}_{R}) ,
\label{eq:HubbardH}
\end{align}
where $\hat{c}_{L(R)\sigma}^{\dag}$ and $\hat{c}_{L(R)\sigma}$ are creation and 
annihilation operators for a spin-$\sigma$ electron on the left(right) site $L(R)$, respectively, and $\hat{n}_{L(R)}=\sum_{\sigma=\uparrow, \downarrow}\hat{c}_{L(R)\sigma}^{\dag} \hat{c}_{L(R)\sigma}$ are the site-occupancy operators.

 The occupation difference $\langle \hat{n}_{L} -\hat{n}_{R}\rangle =
 \Delta n$ represents the dipole in this model, $d=\Delta n$, and is the
 main variable~\cite{FFTAKR13}; the total number of fermions is fixed
 at $N=2$. A static potential difference, $\Delta v^{0}= \sum_\sigma
 (v_{L\sigma}^0 - v_{R \sigma}^0)$, renders the Hubbard dimer
 asymmetric. The total external potential $\Delta v (t)$ is given by $\Delta
 v(t)= \Delta v^{0}+ 2{\mathcal E(t)}$, where the last term represents an electric field that we will tune to induce CT between the sites.  An infinitely long-range  molecule is modelled by $T/U \to 0$: in our calculations, we fix the
 interaction strength to be unity, $U=1$ and make the hopping
 parameter $T$ small,  corresponding to a large separation
 between the sites (equivalent to the strongly correlated limit $U/T \to \infty$).  We
use $\hbar=e=1$ throughout, and all energies are given in units of $U$.

\begin{figure}
\begin{center}
\includegraphics[width=0.5\textwidth]{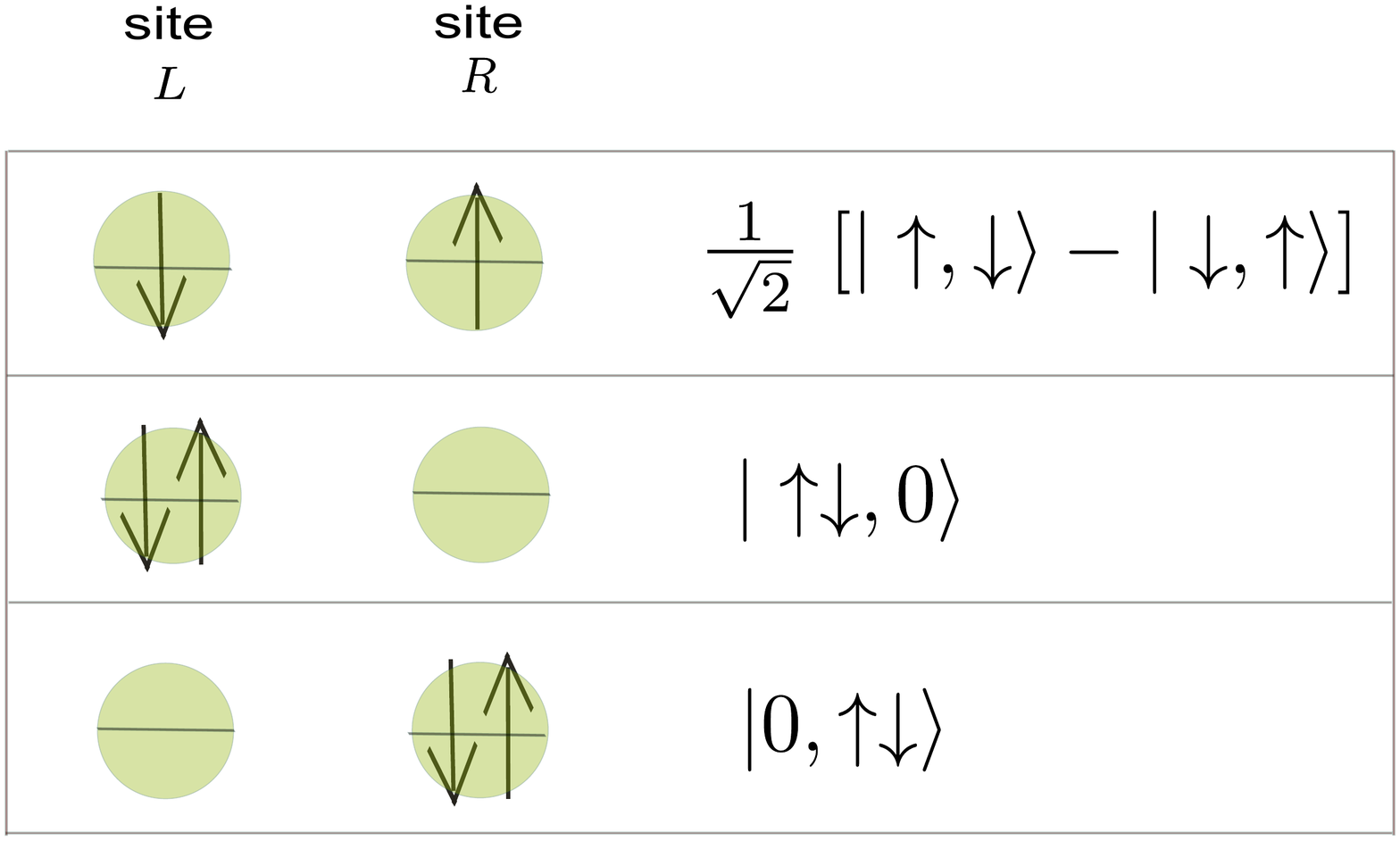}
\caption{The three  states pictured form a complete basis of the singlet sector of the Hubbard dimer.}
\label{fig:Vspace}
\end{center}
\end{figure}
The singlet sector of the two-electron vector space is three-dimensional (depicted in Fig.~\ref{fig:Vspace}),
\bea
|\Psi_1\rangle &=& \frac{1}{\sqrt{2}} \left(\hat{c}_{1\uparrow}^\dag\hat{c}_{2\downarrow}^\dag -\hat{c}_{1\downarrow}^\dag\hat{c}_{2\uparrow}^\dag \right) | 0\rangle=\frac{1}{\sqrt{2}} \ [|\uparrow ,\downarrow \rangle - |\downarrow ,\uparrow \rangle ] \\
|\Psi_2\rangle &=& \hat{c}_{1\uparrow}^\dag\hat{c}_{1\downarrow}^\dag | 0\rangle= |\uparrow \downarrow , 0 \rangle  \\
|\Psi_3\rangle &=& \hat{c}_{2\uparrow}^\dag\hat{c}_{2\downarrow}^\dag | 0 \rangle =  \ |0,\uparrow\downarrow \rangle 
\label{eq:psis}
\eea

For fixed $T/U$ a constrained search search over all gs wavefunctions $|\Psi\rangle = a_1|\Psi_1\rangle +  a_2|\Psi_2\rangle +  a_3|\Psi_3\rangle$ that yield a given $\Delta n$ \cite{L82,L83} can be straightforwardly performed due to the small size of the Hilbert space. This results in the Hohenberg-Kohn (HK) energy functional~\cite{HK64,L82,L83}:
\ben
F_{HK}[\Delta n]=\min_{\Psi \to \Delta n} 
\langle \Psi_{\Delta n} | \hat{T} + \hat{U} | \Psi_{\Delta n} \rangle = E\Hxc[\Delta n] + T_s[\Delta n] \;,
\label{eq:FHK}
\een
where $\hat{T}$ and $\hat{U}$ are the first two terms in Eq.~(\ref{eq:HubbardH}), $T_s[\Delta n]= \min_{\Phi \to \Delta n} \langle \Phi_{\Delta n} | \hat{T} | \Phi_{\Delta n} \rangle$ is the non interacting kinetic energy, and $\Phi$ denotes a single Slater determinant.  $E\Hxc[\Delta n]$ is the HXC energy functional, which must in practise be approximated for real systems, but here for the Hubbard model we can compute it explicitly exactly numerically.
The HK functional $F_{HK}[\Delta n]$ completely determines the gs energy $E_{\rm gs}$,
\ben
E_{\rm gs}= \min_{\Delta n}  \left( F_{\sss HK}[\Delta n] + \frac{\Delta v^0}{2} \Delta n\right)\;.
\label{eq:E0LL}
\een
The gs occupation difference $\Delta n_{\rm gs}$ of all possible asymmetric (and symmetric) Hubbard dimers is determined by $\frac{\partial F_{\sss HK}}{\partial_{\Delta n}}\big|_{\Delta n_{\rm gs}} =\frac{-\Delta v^0}{2}$. 
 
\begin{figure}[ht]
\begin{centering}
\includegraphics[height=0.35\textwidth, clip]{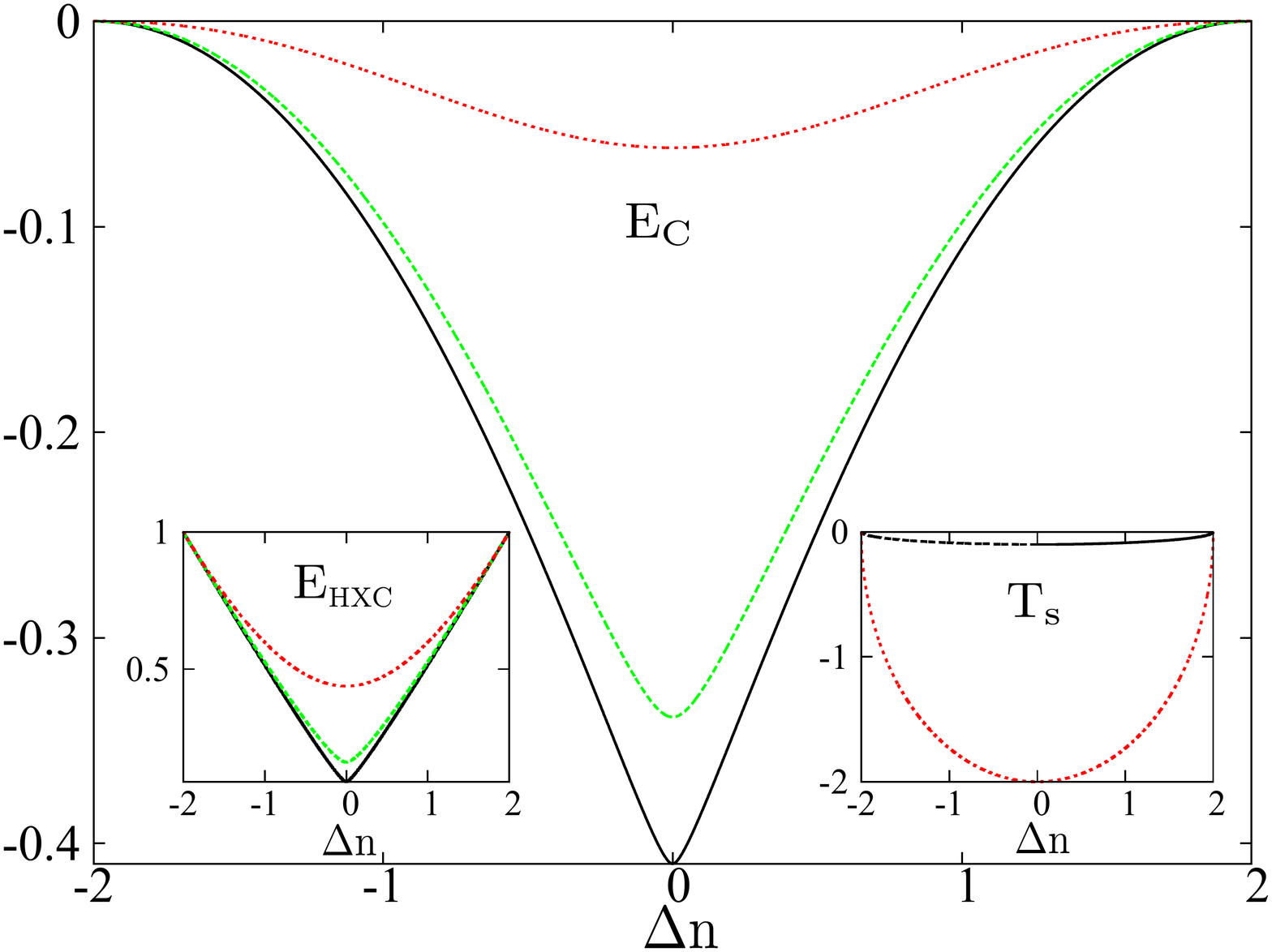}
\caption{ Correlation energy functional $E\c[\Delta n]$ for  $T=1$ (red dotted) $T=0.1$ (green dashed) and $T=0.05$ (black solid).  Insets: $E\Hxc[\Delta n]$ (left) and non-interacting kinetic energy functional $T\s[\Delta n]$ for the same parameters. 
Energies are in units of $U$.
}
\label{fig:Ec}
\end{centering}
\end{figure}

The minimization Eq.~(\ref{eq:FHK}) was carried out in Mathematica; the resulting discrete function  $F_{HK}^j(\Delta n_j)$ 
was fitted and derived using splines to obtain the exact gs HXC potential $\Delta v\Hxc^{\rm gs}[\Delta n]=2 \frac{\partial (F_{HK}[\Delta n] - T_s[\Delta n])}{\partial \Delta n}$ (see Fig.~\ref{fig:Ec}). (The factor $2$ in the right hand side of
Eq.~(\ref{eq:DD}) results from expressing the energy functional in
terms of the variable $\Delta n= n_L - n_R$, namely $\Delta v\c^{\rm gs}[\Delta
  n] = v\c^L[\Delta n] - v\c^R[\Delta n] =\frac{d E\c[\Delta
    n]}{d(\Delta n)} \frac{d \Delta n}{d n_L}-\frac{d E\c[\Delta
    n]}{d(\Delta n)} \frac{d \Delta n}{d n_R}$.)   

\begin{figure}[ht]
\begin{centering}
\includegraphics[height=0.35\textwidth, clip]{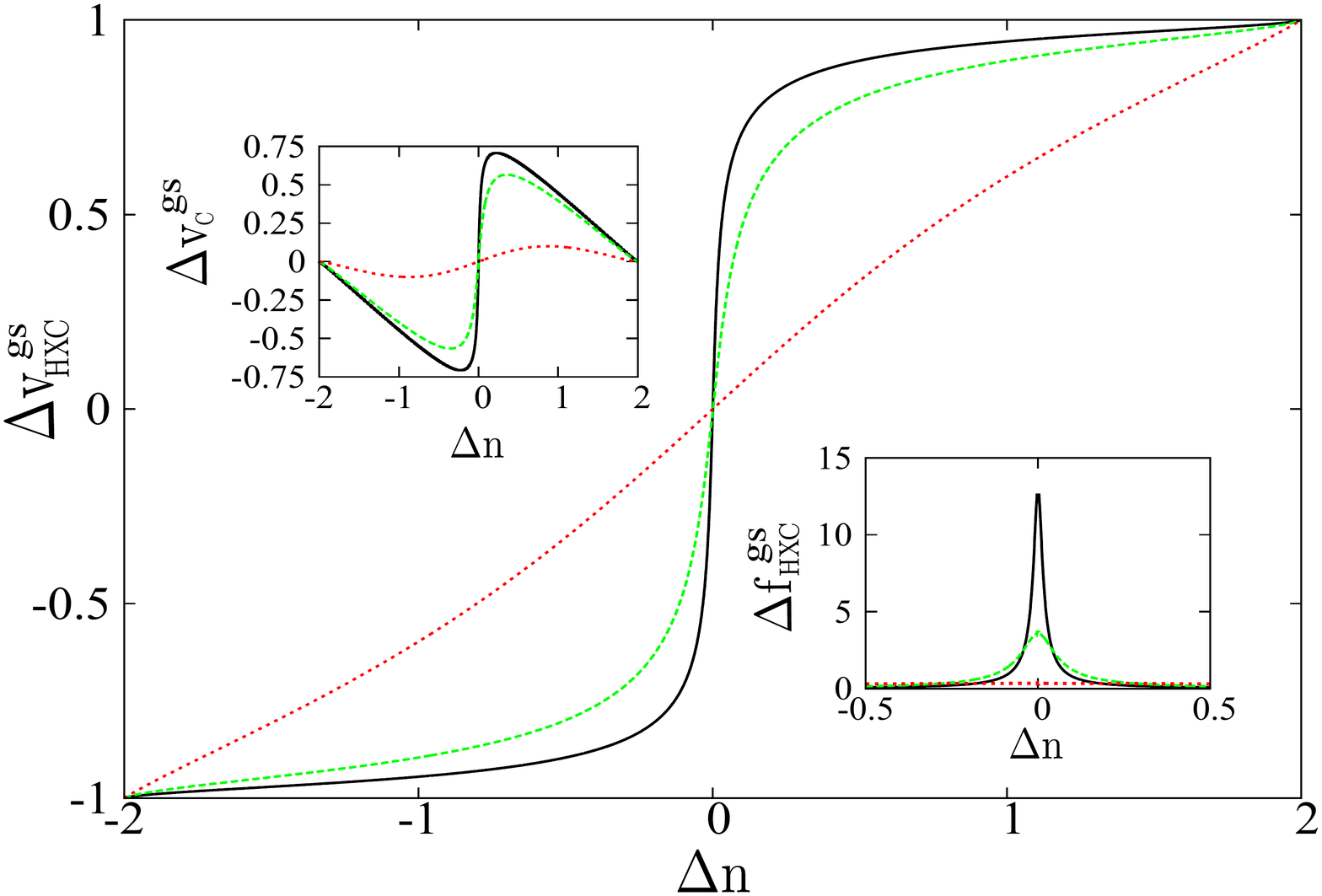}
\caption{Ground-state HXC potential functional $\Delta v\Hxc^{\rm gs}[\Delta n]$  for $T=1$ (red dotted) $T=0.1$ (green dashed) and $T=0.05$ (black solid). Left inset: Ground-state correlation potential functional $\Delta v\c^{\rm gs}[\Delta n]$.
Right inset: $\Delta f\Hxc^{\rm gs}[\Delta n] = \frac{d^2 E\Hxc[\Delta n]}{d(\Delta n)^2}$.
}
\label{fig:Dvc}
\end{centering}
\end{figure}
\vspace{0.5cm}

In Fig.~\ref{fig:Ec} the different components of the
energy as functions of the occupation difference $\Delta n$ for
different  $T$ and fixed Hubbard strength $U=1$ are
plotted.  
 Note that the Hartree-exchange (HX) part of the energy functional is independent of 
 $T$, $E\Hx[\Delta n]=(N^2 +
\Delta n^2)U/8$ (with $N$ being the number of particles)
\cite{CC13}.
In the limit where the two electrons are sitting on the same site the
HXC energy is entirely due to HX and equal to the on-site repulsion
$U$, $E\Hxc [\Delta n =\pm 2]= E\Hx =U$ (see left inset in 
Fig.~\ref{fig:Ec}). The hopping parameter $T$  plays a
role as soon as the electronic density delocalizes.  As shown in right
inset of Fig.~\ref{fig:Ec} the non-interacting kinetic energy
$T\s[\Delta n]$ depends strongly on $T$.  

For
small occupation differences $\Delta n \to 0$ (one electron on each site), in the infinite
separation limit $T/U \to 0$, the correlation energy $E\c[\Delta n]$
develops a discontinuity in its derivative (see Fig.~\ref{fig:Ec}).  This discontinuity manifests
in a step-like function in the correlation potential difference
$\Delta v\c^{\rm gs}[\Delta n]= 2 \frac{\partial \Delta E\c}{\partial \Delta
  n}$ (see left inset Fig.~\ref{fig:Dvc}).
  This feature is related to the derivative discontinuity of the
  isolated 1-electron site for the following reason. The variable
  $\Delta n$ plays the role of the density-variable, as well as
  directly giving the particle number on each site, $n_{\rm L, R} = 1
  \pm \Delta n/2$.  So, in the isolated-site limit $T/U\to 0$, a
  variation $\delta n$ near $\Delta n=0$ can be thought of as
  adding(subtracting) a fraction of charge $\delta n$ to the
  one-fermion site on the left(right): \bea
\label{eq:DD}
\nonumber &\left.2\frac{d E\c[\Delta
    n]}{d(\Delta n)}\right\vert_{\Delta n=0^+} - \left.2\frac{d
  E\c[\Delta n]}{d(\Delta n)}\right\vert_{\Delta n=0^-} =&\\ &\Delta
v\c^{\rm gs}[\Delta n = 0^+] - \Delta v\c^{\rm gs}[\Delta n = 0^-] \equiv 2
\Delta\c^{\rm 1-site}(N=1)\;& 
\eea 
The difference in the correlation potential as one crosses $\Delta n =
0$, therefore coincides with the derivative discontinuity at $N=1$ of one
site; the value of $\Delta\c^{\rm 1-site}(N=1) = U$.
The discontinuity only shows up in the infinite separation limit; if instead the two sites lie closer to each
other they can not be considered as two separated one-electron systems
and thus moving a fraction of electron back or forth represents a
smooth change in the energy.  

In section \ref{sec:cs--cs} we will study CT dynamics between two
closed-shell fragments (cs--cs) by applying a relatively large static
potential difference such that $\Delta n_{\rm gs} \approx 2$. One
electron will be transferred to the other site by turning on a field
resonant with the CT excitation frequency.  Looking at
Figs.~\ref{fig:Ec}-\ref{fig:Dvc} this corresponds to scanning the
densities starting at the outer right region and finishing at the
central region once the CT state (consisting of two now open-shell
sites) is reached. The AE propagation is performed
using the exact gs HXC potential $\Delta v\Hxc^{\rm gs}[\Delta n]$
shown in Fig.~\ref{fig:Dvc}, i.e. assuming that at every time $t$ the
density $\Delta n(t)$ is the gs density of some potential $\Delta v$.
In section \ref{sec:os--os} we study instead CT between two open-shell
fragments (os--os), starting with a gs consisting of two open-shell
sites each with approximately one electron ($\Delta n_{\rm gs} \approx
0$) that evolves to a CT state with $\Delta n \approx 2$; thus
scanning the densities in a ``time-reversed'' way compared to
Sec.~\ref{sec:cs--cs}, moving from the central region in
Figs.~\ref{fig:Ec}-\ref{fig:Dvc} to the outer region.
We have chosen the $\Delta v^{0}$'s such that the CT density of the cs--cs system is close to the gs density of the os--os system, $\Delta
n^{\rm cs--cs}_{CT}\approx \Delta n^{\rm os--os}_{\rm gs}$ and vice-versa, $\Delta
n^{\rm os--os}_{CT}\approx \Delta n^{\rm cs--cs}_{\rm gs}$.

\subsection{Time-dependent Kohn-Sham potential}
\label{sec:tdKS}
The KS Hamiltonian has the form of Eq.~(\ref{eq:HubbardH})
but with $U = 0$ and  $\Delta v(t)$ replaced by the KS potential difference,
\ben
\Delta v_s[\Delta n, \Phi (t_0)](t) = v\Hxc[\Delta n, \Psi (t_0), \Phi (t_0)](t) +  \Delta v(t),
\een
defined such that the  interacting $\Delta n(t)$  is reproduced.
The exact time-dependent KS potential can be found by inversion of the time-dependent KS equations \cite{TDDFTbook} assuming a doubly-occupied singlet state. This yields~\cite{FT12}
\begin{equation}
\Delta v_{s} [\Delta n, \Phi(t=0)] = - \left(\frac{\ddot{\Delta n} +(2T)^2 \ \Delta n}{\sqrt{ (2T)^2 \left(4 - (\Delta n)^2\right) - (\dot{\Delta n})^2}}\right)
\label{eq:vkstddeltan}
\end{equation}
when the KS initial state is the KS gs.
$\Delta n (t)$ is time-dependent non-interacting $V$-representable as long as the denominator in Eq.~(\ref{eq:vkstddeltan}) does not vanish,
\ben 
|\dot{\Delta n}| < 2 T \sqrt{4- (\Delta n)^2}.
\label{eq:V-rep}
\een 
Condition (\ref{eq:V-rep}) fixes an upper bound to the absolute value of the
link-current $|\dot{\Delta n}|$, which can be identified with the sum of currents flowing along links attached to the site.  On a lattice the
maximum link-current depends
on $T$ (see \cite{FT12} and refs. therein).

\section{closed-shell to closed-shell CT } 
\label{sec:cs--cs}
\begin{figure}
\includegraphics[width=0.5\textwidth]{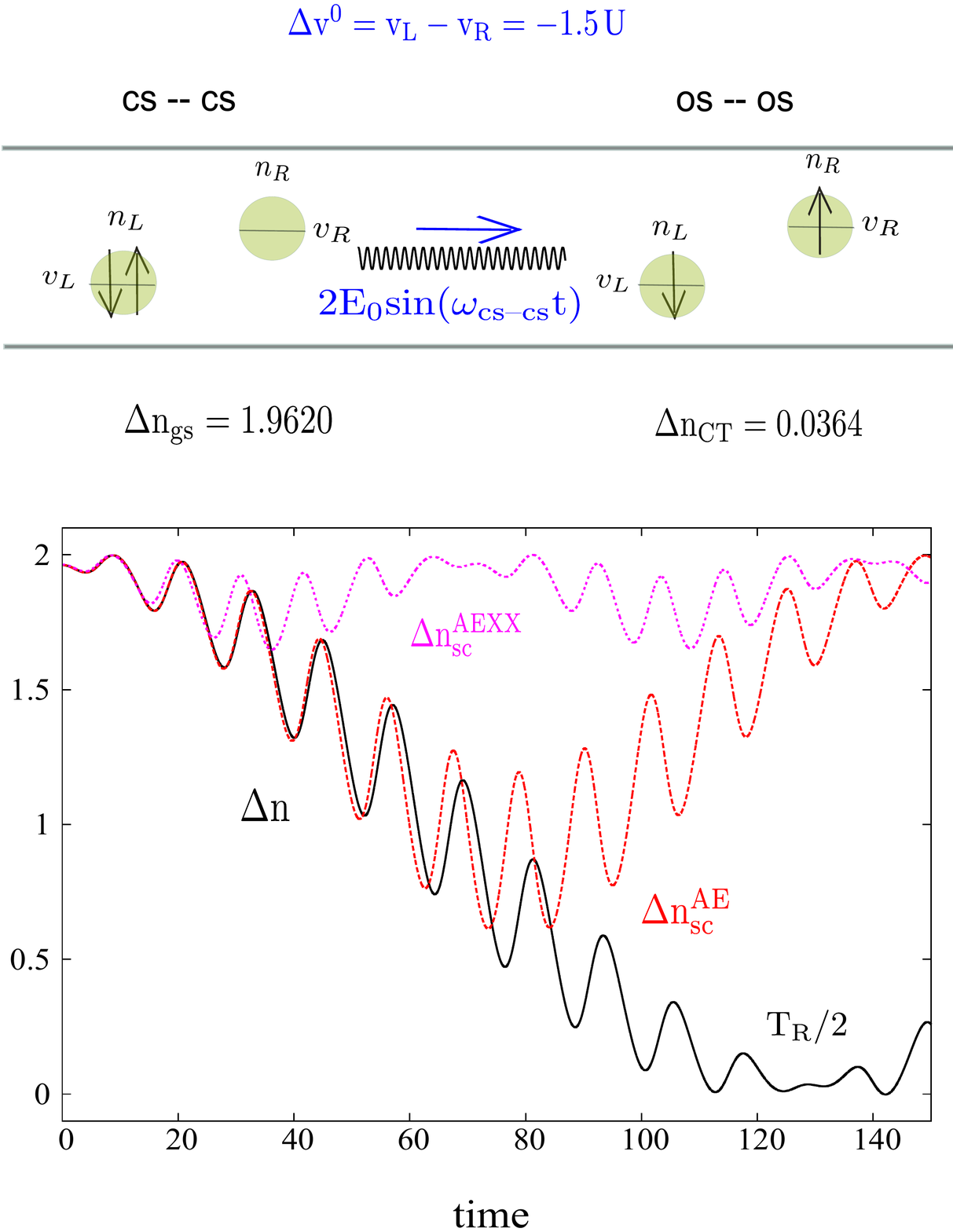}
\caption{Model of CT between two closed-shell fragments at large separation: a large static potential difference $\Delta v^0=-1.5 ~U$ is chosen such that in the gs almost two electrons are sitting on  the left site ($\Delta n_g \approx 2$) and $T/U=0.05$.
In the bottom panel the results of the propagation are shown: exact dipole $\Delta n(t)$ (black solid), self-consistent AE dipole $\Delta n_{sc}^{AE}(t)$ (red dashed) and self-consistent adiabatic EXX dipole $\Delta n_{sc}^{AEXX}(t)$ (pink dotted). Time is given in units of $1/U$, CT state is reached at $T_R/2 \approx 128/U$. 
}
\label{fig:cs-cs}
\end{figure}
To model CT between two closed-shell fragments we choose the static external potential difference in the Hubbard dimer to be $\Delta v^0=-1.5~U$, which results in a gs with almost two fermions sitting on the left site $\Delta n_{\rm gs}=1.9620$ (see top left of Fig.~\ref{fig:cs-cs}). The vector space is built from the three singlet states  introduced in Fig.~\ref{fig:Vspace} and Eqs.~\ref{eq:psis},  and we shall now describe the eigenstates in detail.
The interacting states in this basis were found by first computing the matrix elements of the static Hamiltonian Eq.~\ref{eq:HubbardH} (${\cal E}(t) =0$) in basis Eqs.~\ref{eq:psis} and then diagonalizing the matrix. The calculation was performed in a self-developed code written in the second quantization formalism.
The interacting  gs is predominantly $|\Psi_2 \rangle$:
\ben
\begin{split}
|\Psi_{\rm gs}^{cs-cs}\rangle = &   - 0.13781 \frac{1}{\sqrt{2}}[|\uparrow ,\downarrow \rangle - |\downarrow ,\uparrow \rangle ]  + 0.99045 |\uparrow \downarrow , 0 \rangle\ \\ & + 0.00323 |0,\uparrow\downarrow \rangle,
\label{eq:gs_closed}
\end{split}
\een
while the first excited state is mainly $|\Psi_1 \rangle$, i.e. is a CT excitation with  about one electron on each site,
\ben
\begin{split}
|\Psi_{e1}^{cs-cs}\rangle = &  0.990054 \frac{1}{\sqrt{2}} [|\uparrow ,\downarrow \rangle - |\downarrow ,\uparrow \rangle ]  - 0.13785 |\uparrow \downarrow , 0 \rangle\ \\ & + 0.02809 |0,\uparrow\downarrow \rangle.
\label{eq:CT_closed}
\end{split}
\een
There is a second CT excited state $\Psi_{e_2}^{cs--cs}$,  dominated by $| \Psi_3 \rangle$,
\ben
\begin{split}
|\Psi_{e2}^{cs-cs}\rangle = &  -0.02827 \frac{1}{\sqrt{2}}[|\uparrow ,\downarrow \rangle - |\downarrow ,\uparrow \rangle ]  + 0.000666 |\uparrow \downarrow , 0 \rangle + \\ & 0.9996 |0,\uparrow\downarrow \rangle.
\label{eq:CT2_closed}
\end{split}
\een
The KS states $\Phi$ are obtained from diagonalization of the exact gs KS Hamiltonian, which corresponds to taking $U \to 0$ and $\Delta v = \Delta v^0 + \Delta v\Hxc[\Delta n]$ in Eq.~\ref{eq:HubbardH}, with $\Delta v\Hxc[\Delta n]=2 \frac{\partial E\Hxc[\Delta n]}{\partial \Delta n}$ and $E\Hxc[\Delta n]$ found by constrained search as discussed in section \ref{sec:model}. 
The non-interacting two-electron KS gs is also predominantly $|\Psi_2 \rangle$: 
\ben
\begin{split}
|\Phi_{\rm gs}^{cs-cs}\rangle = &  0.137236 \frac{1}{\sqrt{2}}[|\uparrow ,\downarrow \rangle - |\downarrow ,\uparrow \rangle ]  + 0.99049 |\uparrow \downarrow , 0 \rangle + \\ & 0.00950725 |0,\uparrow\downarrow \rangle,
\label{eq:ks_gs_cs}
\end{split}
\een
the first KS excited state, predominantly $|\Psi_1 \rangle$, corresponds to a single excitation to a CT state:
\ben
\begin{split}
|\Phi_{e1}^{cs-cs}\rangle = & 0.980985 \frac{1}{\sqrt{2}}[|\uparrow ,\downarrow \rangle - |\downarrow ,\uparrow \rangle ]  - 0.137236 |\uparrow \downarrow , 0 \rangle \\ & + 0.137236 |0,\uparrow\downarrow \rangle,
\label{eq:ks_CT_cs}
\end{split}
\een
and the second KS excited state $\Phi_{e_2}^{cs--cs}$ is dominated by $| \Psi_3 \rangle$ and is actually a double excitation:
\ben
\begin{split}
|\Phi_{e2}^{cs-cs}\rangle = &  -0.137236  \frac{1}{\sqrt{2}}[|\uparrow ,\downarrow \rangle - |\downarrow ,\uparrow \rangle ] + 0.0095072 |\uparrow \downarrow , 0 \rangle \\ & + 0.990492 |0,\uparrow\downarrow \rangle. \\ 
\label{eq:ks_e2_cs}
\end{split}
\een
Comparing the KS states with the interacting states Eqs.~(\ref{eq:gs_closed}-\ref{eq:CT2_closed}) we see they are very similar. 

Table~\ref{tab:closed_shell} contains the energies, site-occupation differences $\Delta n=\langle \Psi^{cs-cs}\vert \hat{\Delta n} \vert \Psi^{cs-cs}\rangle$  of the three interacting and KS states enumerated in Eqs.~(\ref{eq:gs_closed}-\ref{eq:ks_e2_cs}),  and the transition matrix elements from the gs to the CT excited states, $d_{gs\to e1(2)} =\langle \Psi_{e1(2)}^{cs-cs}\vert \hat{\Delta n} \vert \Psi_{\rm gs}^{cs-cs}\rangle$ for both interacting and Kohn Sham systems. 
By construction the exact gs HXC functional $E \Hxc$ reproduces the exact gs energy $E_{\rm gs}$ and gs density $\Delta n_{\rm gs}$. All other KS variables shown such as interacting excitation frequencies $\omega\^s$ and transition matrix
elements $d\^s_{gs \to e1(2)}$  have limited physical meaning.  For the case of the cs--cs CT we are studying in this section however, they turn out to be good approximations to the exact quantities,$\omega\^s_{e(2)} = \epsilon\^s_{e(2)} - \epsilon\^s_{\rm gs} \approx \omega$, and
$d_{gs \to e1(2)}\^s = \langle \Phi_{\rm gs} | \hat{\Delta n}| \Phi_{e(2)} \rangle \approx d_{gs \to e1(2)}$ .   
However, in contrast to the
interacting system the non-interacting Kohn-Sham system has equidistant
excitations $\epsilon^s_{e2} - \epsilon^s_{e1}= \epsilon^s_{e1} -
\epsilon^s_{\rm gs} $. That is, the second KS excitation is a pure double-excitation out of the doubly-occupied gs KS orbital; consequently, its dipole transition matrix element is exactly zero. 
In the interacting system, the second excitation has a very small but non-zero transition matrix element (see Table~\ref{tab:closed_shell}). We will choose a field resonant with the first excitation, weak enough that only the ground and first excited interacting states get occupied during the dynamics.



\begin{table}
\caption{Eigenstates, energies, and transition matrix elements for the dimer with $\Delta v^0 = -1.5~U, T=0.05, U=1$. Energies are given in units of $U$. Note that $e1$ and $e2$ are CT excitations for both the interacting and KS systems.}
\begin{center}
\begin{tabular}{|l||c|c|}\hline
    & \multicolumn{2}{|c|}
{$\Delta v=-1.5~U$} 
\\
    & interacting & Kohn-Sham \\
\hline \hline 
$\Delta n_{\rm gs}$      & 1.9620    & 1.9620                   \\
$E_{\rm gs}$      & -0.5098          & -0.5098                       \\
$\epsilon_{\rm gs}$      &     & -0.5152                 \\
\hline \hline
$\Delta n_{e1}$      & 0.0364   & 0.0000                \\
$\epsilon_{e1}$      & 0.0078 & 0.0000                \\
$\omega_{gs \to e1}$      & 0.5177 & 0.5152                \\
$d_{gs \to e1}$      & -0.2733    & -0.2745   \\ 
\hline \hline
$\Delta n_{e2}$ & -1.9980    & -1.9620                 \\
$\epsilon_{e2}$  & 2.5020     & 0.5152   \\
\hline \hline
$\omega_{gs \to e2}$      & 3.0118  & 1.0304                \\
$\omega_{e1 \to e2}$      & 2.4942  &  0.5152                \\
$d_{gs\to e2}$      & -0.0052    &  0.0000  \\
$d_{e1 \to e2}$      & -0.0563    &  -0.2745  \\
 \hline 
\end{tabular}
\label{tab:closed_shell}
\end{center}
\end{table}  

We induce the CT dynamics by turning on  a field resonant with the lowest excitation, $\mathcal{E}(t) = 0.09 \sin(\omega~t)$, with $\omega=\omega_{gs \to e1}=0.5177~U$. 
All propagations were performed using the Crank-Nicholson scheme and a time-step of $0.01/U$.
We evolve the interacting gs in the Hamiltonian of Eq.~\ref{eq:HubbardH} to obtain the exact dipole shown in the lower part of Figure~\ref{fig:cs-cs} for a little over half a Rabi period;
the CT excited state is reached at around $t=128/U$. 
The physics is similar to the real-space CT dynamics in the long-range one-dimensional molecule shown in Figure 4 of
Ref.~\cite{FERM13} (see also Figure 3 in Ref.~\cite{FM14a})
 \footnote{Figure 3 of Ref.~\cite{FM14a}, which was for the Hubbard dimer with a larger asymmmetry, happens to have a closer match with the real-space case.  The resonant frequency was larger, so the dipole oscillates faster in the half-Rabi period and also the transition matrix element $d_{gs \to e1}$ was smaller, making the amplitude of the fast oscillations smaller.} 
and also in the three-dimensional LiCN molecule in Figures 3 and 4 in Ref.~\cite{Nest}.
 Fig.~\ref{fig:cs-cs} shows also the dipole under propagation with the adiabatic exact-exchange (AEXX) approximation, $\Delta v \Hxc^{\rm AEXX}=\Delta v\Hx = 2 \frac{\partial E \Hx[\Delta n]}{\partial \Delta n}=\frac{U}{2} \Delta n(t)$~\cite{CC13}. $\Delta n_{sc}^{\rm AEXX}$ does not show any charge-transfer, resembling the real-space
 AEXX case of Ref.~\cite{FERM13}. Other
 adiabatic approximations were also shown to fail in a
 similar way \cite{FERM13, Nest}.  However in Refs.~\cite{FERM13, Nest} it was not possible to determine
 whether the culprit was the adiabatic approximation itself or the chosen gs
 approximation.
For the Hubbard dimer, with its vastly reduced Hilbert
 space, and the exact HXC potential found by the constrained search (section \ref{sec:model})
we are  able to propagate the KS system with the AE functional; at each
 time-step inserting the instantaneous density $\Delta n^{\rm
   AE}_{sc}$ into the exact gs HXC potential
$\Delta v\Hxc^{\rm gs}[\Delta n^{\rm AE}_{sc}]$ (Fig.~\ref{fig:Dvc}).
The result is $\Delta n^{\rm AE}_{sc}$ on the bottom of Fig.~\ref{fig:cs-cs}:
$\Delta n^{\rm AE}_{sc}$ closely follows the exact density for a
longer time than the AEXX does, but ultimately fails to transfer the
charge. As was concluded in Ref.~\cite{FM14a}, it is essential to have
a memory-dependent functional in order to correctly describe a full
charge transfer.

Ref.~\cite{FM14a} plotted the exact and AE potentials for the case of
cs--cs CT studied there, which illuminated some of the aspects of the dynamics, and
strengthened the comparison with the real-space molecular
case. Although the case studied in Ref.~\cite{FM14a} was for a more
asymmetric dimer ($\Delta v = -2$), resulting in a higher resonant
field frequency and more oscillations over the Rabi period, the essential
observations carry over to the present case, and the potentials follow
similar features to those shown in
Ref.~\cite{FM14a}. 
In particular, (i) the
exact correlation potential drops from its gs value to that of
$-\Delta v^0$ after half a Rabi cycle, such that the total KS
potential $\Delta v\s = \Delta v^0 + \Delta v\Hxc$ goes to zero, equalizing the levels on each site. This exactly mirrors the real-space
case, where a spatial step in the correlation potential in the
intermolecular bonding region develops such that at half-Rabi cycle,
the two atomic levels are aligned with each other, i.e. the step has a
size equal to the difference in the ionization potentials of the
$(N-1)$-electron ions. (ii) The AE correlation potential evaluated on
the {\it exact} density, $\Delta n^{\rm AE}[\Delta n]$,  tracks the ground-state correlation potential
shown in Fig.~\ref{fig:Dvc}, moving from the right inwards to the central region,
gently oscillating around it, in synch with the density. 
As $\Delta n(t)\to 0$ and the CT state is reached it tracks the approaching discontinuity, which in the limit of
$T/U \to 0$, is equal to the one-site
one-fermion derivative discontinuity~\cite{FM14a}, in complete analogy with the infinite-separation limit
of the real-space molecular case~\cite{FERM13}.  
The donor potential is shifted upwards relative to the acceptor by an
amount equal to the derivative discontinuity of the donor, and in both
the real-space and Hubbard cases, this underestimates the shift
provided by the exact correlation potential. (iii) The self-consistent
AE correlation potential, $v\c^{\rm AE}[\Delta n_{sc}](t)$, deviates
from the true potential quite early on.
As a consequence of this,
the two sites remain far from being ``aligned'', forbidding
the possibility that a stable CT state with one electron on each, can be approached in the self-consistent AE propagation.

\begin{figure}
\includegraphics[width=0.5\textwidth]{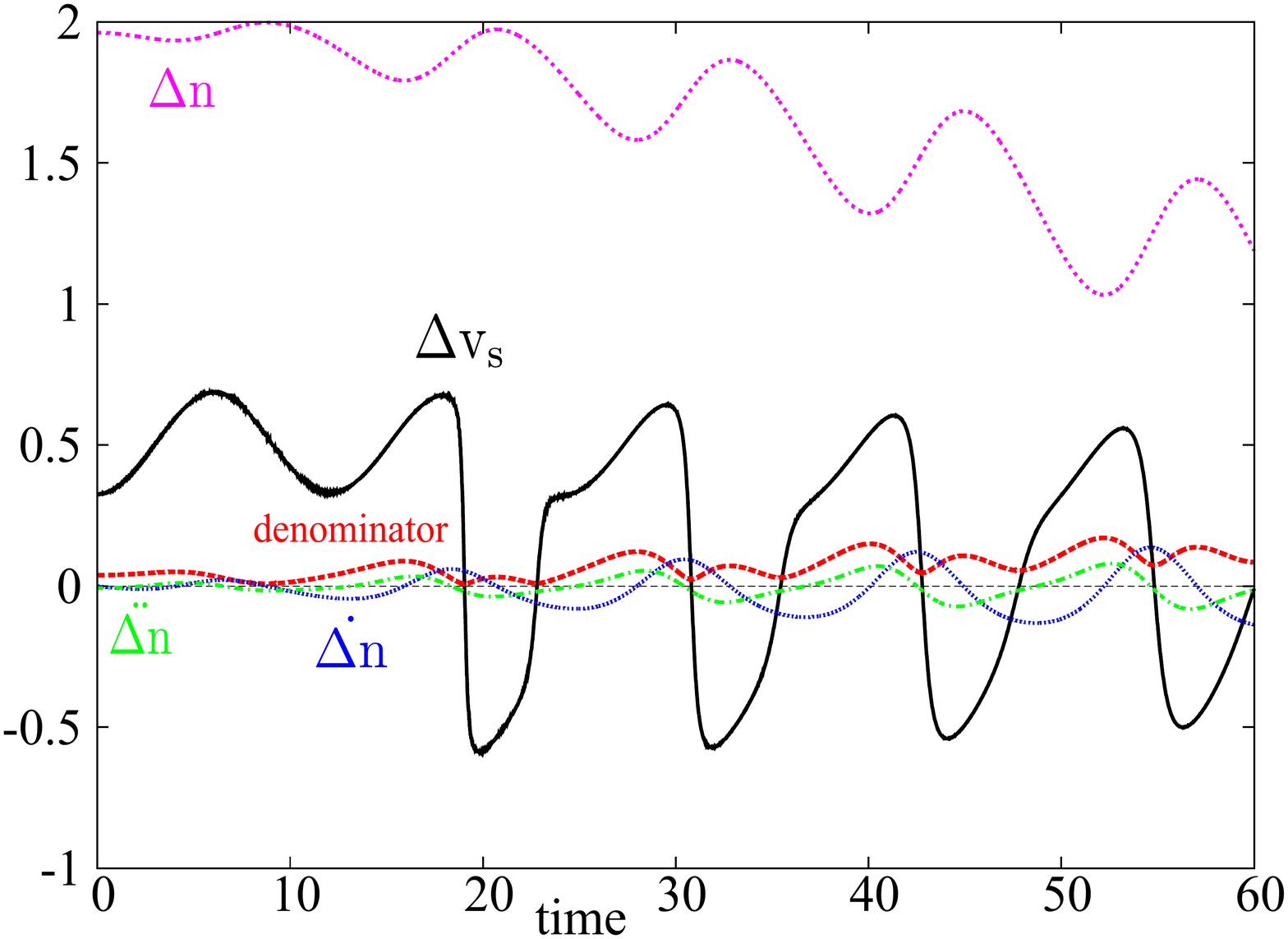}
\caption{$\Delta v^0=-1.5 ~U$ cs--cs CT: KS potential $\Delta v\s$ (black solid), denominator of $\Delta v\s$ (red dashed), link-current $\dot{\Delta n}$ (blue dotted), acceleration $\ddot{\Delta n}$ (green dashed-dotted) and dipole $\Delta n$ (pink dotted).}
\label{fig:v-rep}
\end{figure}

We now turn to one aspect of the exact correlation potential that was
discussed only briefly in Ref.~\cite{FM14a}.  It was found that the
exact correlation potential after a very short time develops 
large oscillations which appear to be related to maintaining
non-interacting v-representability. Since the system begins with
$\Delta n= 1.960$, close to 2, and $T$ is small, the right-hand-side
of Eq.~\ref{eq:V-rep} starts out quite small. The left-hand-side
starts from zero and increases but it does not take a very large
link-current for the two sides of Eq.~\ref{eq:V-rep} to approach each
other, leading to the denominator of $\Delta v\s$ to approach zero and
hence becoming close to violating the non-interacting v-representability
condition.  Figure~\ref{fig:v-rep} shows the early-time behavior of
the KS potential: $\Delta v\s$ first swings sharply to  $-\Delta v\s$
(near time of $19/U$) when the denominator gets very close to
zero, causing the acceleration $\ddot{\Delta n}$ to change direction (smoothly) and a consequent decrease in the current. This moves the
denominator away from zero, escaping the violation of
v-representability. The system oscillates due to the field, and again
the denominator becomes very small at around time of $22/U$, when again the $\Delta v\s$ changes direction, avoiding again the crash into non-v-representability. As time evolves the density transfers, $\Delta n$ moves further from 2, and so larger currents are possible without danger of the v-representability condition being violated. The potential oscillations then become more gentle, as shown in the figure. 

\section{Open-shell to open-shell CT}
\label{sec:os--os}
\begin{figure}
\includegraphics[width=0.5\textwidth]{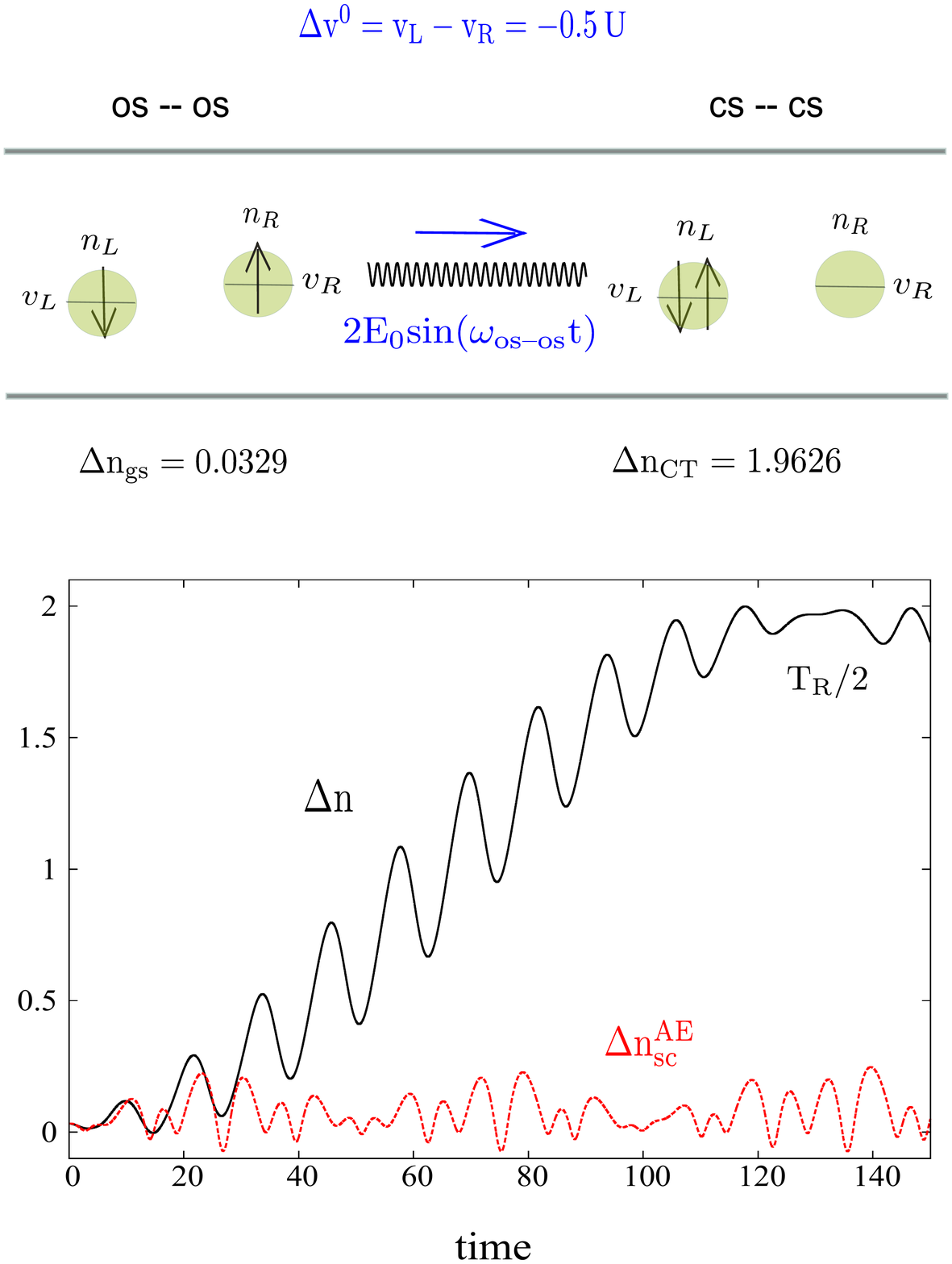}
\caption{Model for CT between two open-shell fragments at large separation:a small static potential difference $\Delta v^0=-0.5 ~U$, is chosen such that the gs is close to  homogenous ($\Delta n_g \approx 0$), and again $T/U=0.05$. In the bottom panel the results of the propagation are shown: exact dipole $\Delta n(t)$  (black solid), self-consistent AE dipole $\Delta n_{sc}^{AE}(t)$ (red dashed).
 Time is given in units of $1/U$; the CT state is reached at $T_R/2 \approx 129/U$. (The AEXX dipole is not included, because of difficulties in converging to a stable os--os gs for these parameters). }
\label{fig:os--os}
\end{figure}

To study CT between two open-shell fragments we choose the static
external potential difference to be $\Delta v^0=-0.5~U$, which results
in a gs with about one electron on each site, $\Delta n= 0.0329$ (see
Fig.~\ref{fig:os--os}). The value of $\Delta v^0$ has been chosen such
that $\Delta n_{\rm gs}^{\Delta v =-0.5} \approx \Delta n_{CT}^{\Delta v =
  -1.5}$ of the cs--cs case in the previous section, and the CT
excitation in the present os--os case has $\Delta n_{CT}^{\Delta v
  =-0.5} \approx \Delta n_{\rm gs}^{\Delta v = -1.5}$ of the os--os
case. (Compare Table~\ref{tab:closed_shell} and
Table\ref{tab:open_shell}).
The gs of this problem is dominated by $|\Psi_1\rangle$:
\ben
\begin{split}
|\Psi_{\rm gs}^{os-os}\rangle = &  0.989568 \frac{1}{\sqrt{2}}[|\uparrow ,\downarrow \rangle - |\downarrow ,\uparrow \rangle ]  + 0.136387 |\uparrow \downarrow , 0 \rangle \\ & + 0.04625 |0,\uparrow\downarrow \rangle ,
\label{eq:gs_open}
\end{split}
\een
while the CT excited state is mainly $|\Psi_2\rangle$:
\ben
\begin{split}
|\Psi_{e1}^{os-os}\rangle = &  -0.13608 \frac{1}{\sqrt{2}} [|\uparrow ,\downarrow \rangle - |\downarrow ,\uparrow \rangle ]  + 0.99065 |\uparrow \downarrow , 0 \rangle \\ & -0.0097168 |0,\uparrow\downarrow \rangle,
\label{eq:CT_open}
\end{split}
\een
with almost two electrons on the left site.
Again there is a second CT state, with much smaller gs dipole transition matrix element $d_{g \to CT_2} << d_{g \to CT}$ (see Table~\ref{tab:open_shell}), and close to both electrons on the right ($|\Psi_3\rangle$):
\ben
\begin{split}
|\Psi_{e2}^{os-os}\rangle = &  -0.0047131 \frac{1}{\sqrt{2}}[|\uparrow ,\downarrow \rangle - |\downarrow ,\uparrow \rangle ]  + 0.0033222 |\uparrow \downarrow , 0 \rangle \\ & + 0.99888 |0,\uparrow\downarrow \rangle.
\label{eq:CT2_open}
\end{split}
\een
The ground and excited KS states have a very different form: instead of the predominantly Heitler-London-like nature of the interacting gs, the ground-KS state is a SSD of the form:
\ben
\begin{split}
|\Phi_{\rm gs}^{os-os}\rangle = & 0.707011 \frac{1}{\sqrt{2}} [|\uparrow ,\downarrow \rangle - |\downarrow ,\uparrow \rangle ]  + 0.50823 |\uparrow \downarrow , 0 \rangle\ \\ & + 0.49177 |0,\uparrow\downarrow \rangle.
\label{eq:ks_gs_open}
\end{split}
\een
This is quite analogous to the real-space molecular case where the KS
state is a doubly-occupied bonding orbital, a single Slater
determinant, while the interacting is of Heitler-London form in the
infinite separation limit, requiring minimally two determinants to
describe. The KS excitations are also similar to the real-space case: the first KS excitation is not a CT state, but rather a single-excitation to the antibonding state, and second KS excitation is a double-excitation to the antibonding state:
\ben
\begin{split}
|\Phi_{e1}^{os-os}\rangle = &  -0.016462\frac{1}{\sqrt{2}} [|\uparrow ,\downarrow \rangle - |\downarrow ,\uparrow \rangle ]  + 0.707011 |\uparrow \downarrow , 0 \rangle \\ & -0.707011 |0,\uparrow\downarrow \rangle
\label{eq:ks_e1_open}
\end{split}
\een
\ben
\begin{split}
|\Phi_{e2}^{os-os}\rangle = &  -0.707011 \frac{1}{\sqrt{2}} [|\uparrow ,\downarrow \rangle - |\downarrow ,\uparrow \rangle ]  + 0.49177 |\uparrow \downarrow , 0 \rangle \\ & + 0.50823 |0,\uparrow\downarrow \rangle.
\label{eq:ks_e2_open}
\end{split}
\een
As a consequence the KS excitation energies become very small as $T/U \to 0$ in contrast with the true energies, just as in the infinite-separation limit of the real-space case, and it can also be understood from the
above that the transition matrix elements are large in the KS case while small in the true case (Table~\ref{tab:open_shell}).


\begin{table}
\caption{Eigenstates, energies, and transition matrix elements for the dimer with $\Delta v^0 = -0.5~T, T=0.05, U=1$. Energies are given in units of $U$. Note that $e1$ and $e2$ are CT excitations for the interacting system.}
\begin{center}
\begin{tabular}{|l||c|c|}\hline
    & \multicolumn{2}{|c|}{$\Delta v=-0.5~T$} \\
    & interacting & Kohn-Sham \\
\hline \hline 
$\Delta n_{\rm gs}$      & 0.0329    & 0.0329                   \\
$E_{\rm gs}$      & -0.0131          & -0.0131                       \\
$\epsilon_{\rm gs}$      &    & -0.1000                 \\
\hline \hline
$\Delta n_{e1}$      & 1.9626   &  0.000                \\
$\epsilon_{e1}$      & 0.5097 & 0.000                \\
$\omega_{gs \to e1}$      & 0.5228 & 0.1000                \\
$d_{gs \to e1}$      & 0.2711    & 1.4140   \\ 
\hline \hline
$\Delta n_{e2}$ & -1.9955    & -0.0329                 \\
$\epsilon_{e2}$  & 1.5033     & 0.1000   \\
\hline \hline
$\omega_{gs \to e2}$      & 1.5164  & 0.2000                \\
$\omega_{e1 \to e2}$      & 0.9936  & 0.1000                \\
$d_{gs \to e2}$      & -0.0915    &  0.0000  \\
$d_{e1 \to e2}$      & 0.0260    &  1.4140  \\
 \hline 
\end{tabular}
\label{tab:open_shell}
\end{center}
\end{table}

We now turn to the dynamics, taking $\mathcal{E}(t) = 0.09 \sin(\omega t)$, resonant with the lowest CT excitation resonance, $\omega = \omega_{gs  \to e1} = 0.5228 ~U$, and compare the exact and AE dipoles, as in the previous section. 
Due to the present choice of parameters, the exact dipole
looks almost exactly like a mirror image of the cs--cs case in the
previous section, i.e. the dipole dynamics in the os--os case resembles that of the cs--cs case starting at the CT state.
However the AE dipole does not at all. 
The AE dipole for the os--os case fails badly even after a
very short time, as shown in Fig.~\ref{fig:os--os}; 
for all times one electron more or less
remains on each site during the AE propagation, while in the exact
propagation, (almost) one electron transfers from the left to the
right site. 

The exact and AE potentials are similar to the os--os case studied in
Ref.~\cite{FM14a} for a slightly smaller asymmetry ($\Delta v^0 =
-0.4$ in Ref.~\cite{FM14a}).  The essential features are as follows.
The exact correlation potential has the same property as in the
real-space case: it starts with a value to exactly cancel the
asymmetry in the external potential, such that the KS potential sees
the two sites aligned. In the real-space case, the HXC gs potential of
a long-ranged heteroatomic diatomic molecule has a step in the bonding
region that aligns the highest occupied orbital energies on each
atom~\cite{P85b,GB96,TMM09}. In both real-space and Hubbard dimer
cases, this is a ground-state correlation effect.  Then, as the charge
transfers, the relative shift in the correlation potential between the
sites oscillates on the optical scale while dropping to the value
predicted by subtracting the external potential from
Eq.~(\ref{eq:vkstddeltan}), putting $\dot\Delta n = \ddot\Delta n=0$,
when the CT excited state is reached.  
 As for $\Delta
v\c^{\rm AE}[\Delta n](t)$, it tracks $\Delta v\c^{\rm gs}[\Delta
  n(t)]$ of Fig.~\ref{fig:Dvc} moving from near the center out to the
right; with gentle oscillations reflecting the oscillations in $\Delta
n(t)$. Again we note that its value at the CT excited state is the
correlation potential of a gs of density $\Delta n = 1.9626$ as
opposed to the exact correlation potential which is that for an
excited-state of the same density.  On the other hand, the
self-consistent AE potential, although it starts correctly (both KS
and interacting initial states being ground-states) and captures the
relative shift between the sites, quite quickly deviates from the
exact. This is because the energy of the lowest excitation of the KS
system is very close to the gs (see Table~\ref{tab:open_shell}); the
system becomes increasingly degenerate as $T/U \to 0$. This is quite
in contrast to the true interacting system which has Heitler-London
form in the gs and a finite gap $\omega_{gs \to e1}$.  To open the
vanishing KS gap $\omega\^s_{gs \to e1}$ strong non-adiabaticity is
required in the linear response kernel~\cite{MT06,EGCM11}; the reason
is that the double excitation is nearly degenerate with the single
excitation and thus critical to incorporate. Given that at short times the dynamics is close to the linear response regime, this might explain why the adiabatic propagation of the os--os system fails so early. 
Given the analogous structure of the states for a real-space molecule composed of open-shell fragments, we expect that also in real space a self-consistent AE calculation will lead to a very poor dipole.


\section{Linear response formula}
\label{sec:AEres}
The results above show the failure of AE TDDFT to yield accurate CT
dynamics in both the case when the CT is between closed-shell
sites and when it is between open-shell sites. In the former case we
noted that the KS excitation frequencies were close to the exact,
while in the latter case they were significantly different. We now ask
what the AE TDDFT frequencies are in each case, i.e. when an AE kernel
is used in linear response, to check whether there is an indication of
the bad CT dynamics of the AE in its predicted excitation energies.

First we derive a general expression for the TDDFT excitation energies  of the Hubbard dimer, based on the dipole-dipole response function:
\ben
\chi_{\hat{\Delta n}, \hat{\Delta n}}(\omega)= \frac{d \Delta n}{d (\Delta v/2)}\bigg|_{\Delta n_{\rm gs}},
\label{eq:chi}
\een
where the factor of $1/2$ comes from the fact that the potential difference couples to the dipole operator $\hat{\Delta n}=\hat{n}_L - \hat{n}_R$ with a factor of $1/2$ in Hamiltonian in Eq.~(\ref{eq:HubbardH}). 
From the relation $\Delta v\s = \Delta v +\Delta v\Hxc$, we then find a Dyson-like equation relating $\chi_{\hat{\Delta n}, \hat{\Delta n}}(\omega)$ to  
the KS linear response function and the kernel:
\ben
\chi^{-1}_{\hat{\Delta n}, \hat{\Delta n}}(\omega) = \chi^{-1}_{s, \hat{\Delta n}, \hat{\Delta n}}(\omega) - \Delta f\Hxc(\omega),
\label{eq:dyson}
\een 
where $\Delta f\Hxc [\Delta n]=d(\Delta v \Hxc[\Delta n]/2)/ d (\Delta n)$.
In the KS linear response function, 
\ben
\chi_{s, \hat{\Delta n}, \hat{\Delta n}}(\omega) = \sum_j \frac{\langle \Phi_{\rm gs} | \hat{\Delta n} | \Phi_{ej}\rangle \langle \Phi_{ej} | \hat{\Delta n} | \Phi_{\rm gs}\rangle }{\omega - \omega_{gs \to e_j} + i\eta} + c.c.
\label{eq:chi_s}
\een 
there is only one term in the sum, since only one excitation contributes, that due to the KS single-excitation $e1$,  as the double-excitation $e2$ yields a zero numerator. 
At a true excitation, $\chi_{\hat{\Delta n}, \hat{\Delta n}}(\omega)$ has a pole in $\omega$, and 
 $\chi^{-1}_{\hat{\Delta n}, \hat{\Delta n}}(\omega)$ vanishes. 
So putting the right-hand-side of Eq.~\ref{eq:dyson} to zero, we obtain the  excitation frequencies of the interacting system from:
\ben
\omega^2 = \omega\s^2 + 2 \omega\s|d^s_{\rm gs \to \rm e1}|^2 \Delta f\Hxc[\Delta n](\omega),
\label{eq:LR}
\een 
where $\omega\s$ is the KS eigenvalue difference $\omega\s=\omega\^s_{gs \to e1}$. 
Eq.~(\ref{eq:LR}) has the same form as the ``small matrix approximation'' of the real-space TDDFT linear response equations~\cite{TDDFTbook} except for a factor of $2$, again due to the
use of $\Delta n$ as main variable.  But an important difference is
that Eq.~(\ref{eq:LR}) is exact, since there is only one KS single-excitation in the Hubbard dimer. 
The correction to the bare KS eigenvalue difference (second term in
Eq.~\ref{eq:LR}) is most significant for os--os ground states, because,
as discussed at the end of section \ref{sec:os--os}, the exact
resonant frequency of the interacting os--os system is finite, while
the resonant frequency of the KS system is very small (bonding --
antibonding transition). As a consequence the exact,
frequency-dependent kernel $\Delta f\Hxc[\Delta n](\omega)$ must be very
large in the os--os case.  On the other hand, if we consider the exact
gs HXC kernel (shown in inset of Fig.~\ref{fig:Dvc}), that yields the TDDFT frequency in the AE approximation, it also becomes large around $\Delta n=0$ (see the inset of Figure~\ref{fig:Dvc}).
That is, 
\ben 
\Delta f\Hxc^{\rm AE}=\frac{d (\Delta v \Hxc^{\rm gs}/2 )}{d \Delta n}\bigg|_{\Delta n_{\rm gs}} = \frac{d^2 E\Hxc}{d \Delta n^2} \bigg|_{\Delta n_{\rm gs}}\;
\label{eq:fhxcAE}
\een
has
 a sharp peaked
structure at $\Delta n = 0$. In the limit that $T/U \to 0$, it becomes proportional
to a $\delta$-function.  This divergence of the static kernel is
consistent with what is found for real os--os molecules at large separation, 
Refs.~\cite{GGGB00,MT06}.


Using Eq.~(\ref{eq:fhxcAE}) in Eq.~(\ref{eq:LR}) gives the AE resonant frequency $\omega^{AE}$. 
For the $\Delta v^0=-0.5~U$ os--os CT of section \ref{sec:os--os} we find the  AE resonance $\omega^{\rm AE}_{os--os}=1.1681 ~U$ overestimates the physical resonance $\omega_{os--os}=0.5228~U$ significantly. There is a large non-adiabatic correction to the static kernel in this case. 
On the other hand, for the cs--cs CT of section~\ref{sec:cs--cs}, the
bare KS eigenvalue difference is already a good approximation to the
true resonance (see table~\ref{tab:closed_shell}), and the correction due
to $\Delta f\Hxc^{\rm AE}$ brings the AE resonance even closer, 
$\omega^{\rm AE}_{cs--cs}=0.5187~U$, only $0.001~U$ away from the true exact
resonance. 
This is consistent with our finding that for short times, the AE cs--cs dipole followed the exact one closely, while the AE
os--os one did not (at short enough times the system responds in a linear way). 
Similarly in Ref.~\cite{Nest} it was shown that despite good LR spectra (Figure 5), the time-resolved CT within LiCN molecule was not predicted by any of the approximate adiabatic functionals tested.   
The failure of AE in the cs--cs case at later times  is not surprising  given the fully non-linear nature of the CT dynamics.

These findings  are analogous to the real-space case: here CT excitation
  energies of a long-range molecule composed of closed-shell
fragments can be well-captured by an adiabatic
approximation~(e.g. Ref.~\cite{GB04b}), but the non-linear process of fully
time-resolved CT, requires a non-adiabatic approximation.  When the
molecule consists of two open-shell fragments, non-adiabaticity is
essential even in the linear response regime~\cite{MT06}.




\section{Conclusions and Outlook}
\label{sec:concl}
The Hubbard dimer with small $T/U$ parameters is useful for studying
real-time CT dynamics in a long-range molecule. Due to its small
Hilbert space much can be done numerically exactly or
  even analytically, so enabling a thorough study of the performance
of the adiabatic approximation in TDDFT, which can not be easily
studied in real-space.  In particular, we examined here the
performance of the AE propagation to describe
time-resolved CT dynamics. Although previous work on
real-space molecules has shown that the usual adiabatic
approximations perform poorly~\cite{Nest}, whether this is largely
due to the choice of gs functional or to the adiabatic approximation
itself was not known. Ref.~\cite{FERM13} showed that the AE approximation when evaluated on the {\it
  exact} density, yields a step structure known to be important
in CT dynamics. This AE step has exactly the right size 
in the case of CT between  two open-shell atoms, where the step appears in the initial potential, 
but the wrong step-size for CT between two closed-shell atoms when the step appears in the final CT excited state. 
By propagating the
Hubbard dimer self-consistently with the AE approximation, a numerically very
challenging task in real-space, we were able to show that the AE
approximation qualitatively fails to decribe
  time-resolved CT dynamics.  In the case of  CT between
open-shell fragments AE fails very early, and actually does not transfer
any charge.
In the case of CT between closed-shells
the collapse of the adiabatic
approximation shows up later in the dynamics: the AE dipole follows
the exact one for a significant part of the Rabi cycle, but it drops
back to its initial value way before the physical system has reached
the CT state. 
One may think that the failure is due to the AE resonant
frequency being detuned from the exact one, but for the cs--cs case
the AE resonance is actually very close to the exact resonance!
Clearly memory effects are essential to describe time-resolved CT.

In both the cs--cs CT and the os--os CT, the form of the interacting
state undergoes a fundamental change: in the cs--cs case, from
approximately a single-Slater determinant initially to a double-Slater
determinant of Heitler-London type in the CT state, while the reverse
occurs for the os--os case. The KS state however remains
a single Slater determinant
throughout (a doubly-occupied orbital singlet state). In a sense, this
is the underlying reason for the development (or loss) of the step
structure in the exact potential in real-space, reflected in the
Hubbard model by the realignment of the two sites, signifying strong
correlation. An AE approximation does capture this strong correlation
effect perfectly when it occurs in the gs, but our work here
shows it cannot propagate well. In the cs--cs case, the AE potential was ultimately unable to develop the shift needed for the CT state. In the os--os case it begins with the correct shift but the
near-degeneracy in the KS system meant that even as soon as we begin to evolve away from the gs, the AE approximation fails.
The main features of the exact time-dependent HXC potential and the exact
gs potential are analogous to the real-space case, in particular the
relative shifts between donor and acceptor and the relation with the
derivative discontinuity. This shift appears as an intermolecular step in the real-space case, but we show here that an 'adiabatic step' is not enough to model the dynamics: the results here suggest that its nonlocal dependence on both space and time must be modelled to yield accurate CT dynamics in molecules. 

 Of course there are many aspects of a real CT within a molecule that
 cannot be modeled by a two-site lattice, nevertheless we stress here
 that even for such a simple model relevant physics of the electronic
 process is missed if an adiabatic approximation is used.  The impact
 of the step structure is likely to be dampened by the effect of many
 electrons, three-dimensions, coupling to ionic motion, etc, but there
 is no reason to believe that the shortcomings of the adiabatic
 approximation to describe time-resolved long-range CT will completely
 disappear when more complexity is added to the model.

\begin{acknowledgments} 
We gratefully acknowledge financial support from the National
Science Foundation CHE-1152784 (NTM) and US Department of Energy Office
of Basic Energy Sciences, Division of Chemical Sciences, Geosciences
and Biosciences under Award DE-SC0008623  (JIF).
\end{acknowledgments}

\end{document}